\documentclass[12pt]{iopart}

\usepackage{iopams,graphicx,cite}

\begin{document}

\title{Refractive index of a transparent liquid measured with a concave mirror}
\author{Amitabh Joshi$^1$ and Juan D Serna$^2$}
\address{$^1$ Department of Physics, Eastern Illinois University,
              Charleston, IL 61920, USA}
\address{$^2$ School of Mathematical and Natural Sciences,
              University of Arkansas at Monticello, Monticello, AR 71656, USA}
\ead{\mailto{ajoshi@eiu.edu}, \mailto{serna@uamont.edu}}

\begin{abstract} 
This paper describes the spherical concave mirror method for measuring the index
of refraction of transparent liquids. We derived the refractive index equation
using Snell's law and the small-angle approximation. We also verified the
validity of this method using the traditional spherical mirror and thin-lens
Gaussian equations.
\end{abstract}



\section*{Introduction}

Measuring the refractive index $n$ of a substance or medium is part of every
introductory physics lab. Various approaches to determine this index have been
developed over the years based on the different ways light reflects and
transmits in the medium.

With the introduction of lasers in basic physics courses, a number of these
methods have become accessible to the undergraduate physics laboratory. Several
of these techniques are highly accurate and use specific optical equipment, like
spectrometers, interferometers, or microscopes. In the laboratory, the method
instructors choose to measure the refractive index of a liquid depends on
different factors, such as the optical properties of the substance they want to
study, or simply the resources available at hand, that in many cases, are very
limited.

Among all those different methods, the spherical concave mirror filled with a
liquid makes itself an excellent alternative to measure the refractive index of
the liquid, specially when no fancy apparatuses are available, and the accuracy
of the measurements is not critical.

In this paper, we would like to present a simple geometrical derivation of the
refractive index of a transparent liquid that is obtained using a spherical
concave mirror. This derivation relies mostly on Snell's law and the small-angle
approximation. In addition, we also use Gaussian optics equations for mirrors
and thin-lenses to verify the validity of the refractive index equation.

This method is based on the measurements of the actual and apparent position of
the centre of curvature of the mirror, when it is empty and filled with a
liquid, respectively. The laws of reflection and refraction are essential to
understand the physics behind this method. Students measuring indices of
refraction of liquids using this technique are introduced gradually with the
concepts of reflection and refraction of light, Snell's law, image formation by
spherical mirrors, and Gaussian optics.

\section*{Spherical mirrors}

\begin{figure}[h]
\centering
\includegraphics[clip=true,scale=0.75]{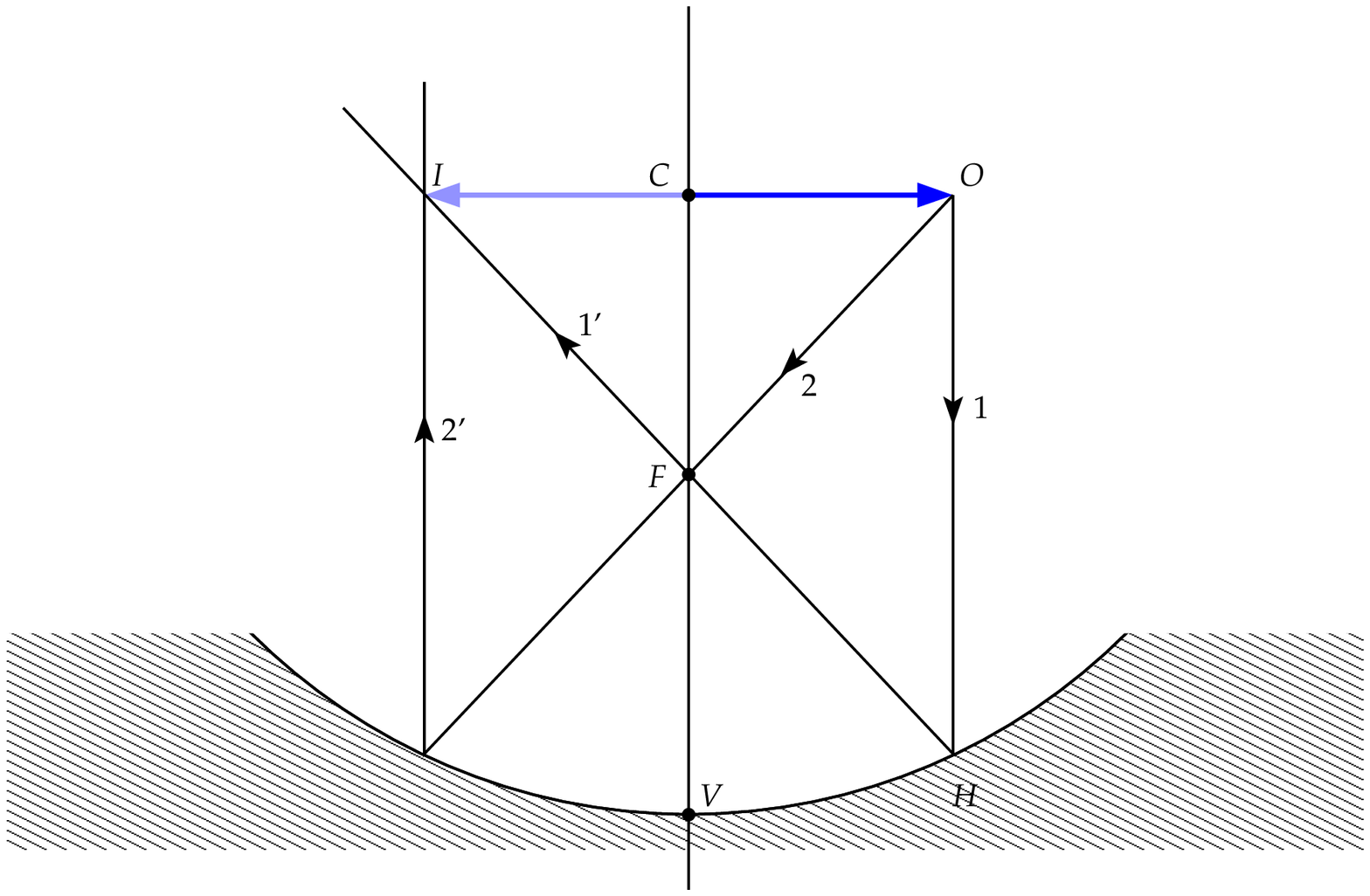}
\caption{\label{Fig:01}Positions of the focal point and centre of curvature of a
spherical concave mirror. Rays 1 and 2, and their respective reflections $1'$
and $2'$ in the mirror, determine the location of the image.}
\end{figure}

Let us first consider the spherical concave mirror shown in~\fref{Fig:01}. The
\textit{optical axis} is the radial line through the centre of the mirror that
intersects its surface at the vertex point $V$. Some relevant points on the
optical axis are the \textit{centre of curvature} $C$, and the \textit{focal
point} $F$. The centre of curvature coincides with the centre of the sphere of
which the mirror forms a section. At the focal point, rays parallel to the
optical axis and incident on the concave mirror, intersect after being reflected
by the mirror's surface~\cite{Wilson}.
The distance $CV$ between the centre of curvature and the vertex is equal to the
radius of the sphere and is called the \textit{radius of curvature} $R$.
Similarly, the distance $FV$ between the focal point of the mirror and the
vertex is known as the \textit{focal length} $f$.

In general, when the rays are close to the optical axis---that is, for the
\emph{small-angle approximation}, the focal length can be shown to be half of
the radius of curvature~\cite{Hecht}:
\begin{equation}\label{Eq:f}
  f = \frac{R}{2}.
\end{equation}

The location and nature of the image formed by a spherical mirror can be
determined by graphical ray-tracing techniques~\cite{Pedrotti}. To find the
conjugate image point $I$ of an object point $O$ located at the centre of
curvature $C$, the paths of any two rays leaving $O$ are
sufficient~\cite{Suppapittayaporn}. We first use the so-called \textit{parallel
ray} $1$ that is incident along a path parallel to the optical axis, strikes the
mirror at point $H$, and is reflected through the focal point $F$ as ray $1'$.
Next, we use the so-called \textit{focal ray} $2$ that passes through the focal
point and is reflected parallel to the optical axis as ray $2'$. The image point
$I$ is formed where the two rays $1'$ and $2'$ intersect. This image is real,
inverted, located at the centre of curvature $C$, and has the same size of the
object, as shown in~\fref{Fig:01}.

\section*{The experiment}

Measuring the refractive index $n$ of a transparent liquid, like water, using a
spherical concave mirror is based on locating the \emph{actual} and
\emph{apparent} centres of curvature of the mirror, when it is empty and filled
with a thin layer of water, respectively.

\begin{figure}[h]
\centering
\includegraphics[clip=true,scale=0.70]{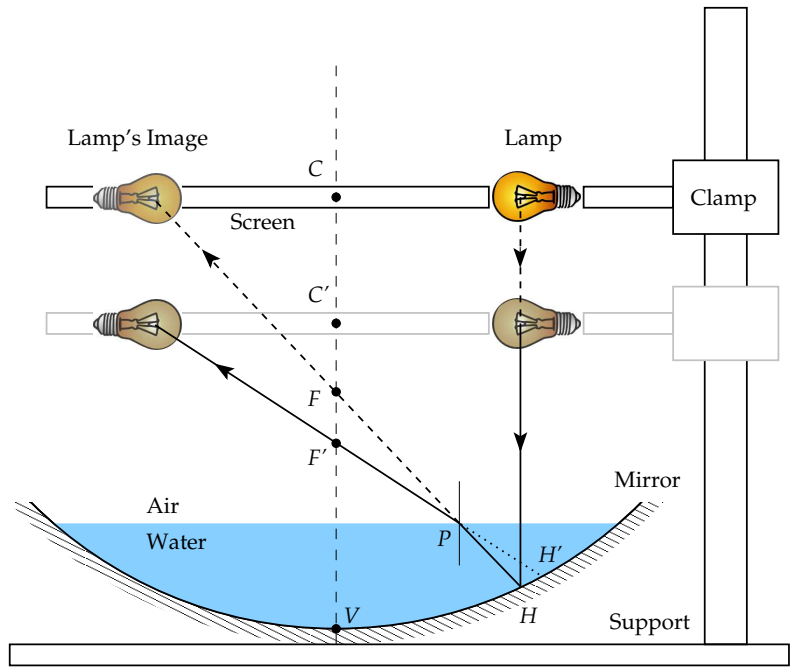}
\caption{\label{Fig:02}Schematics of the spherical concave mirror experiment
for measuring the refractive index of water. The support holding the lamp is
moved up and down until a sharp image of the lamp is formed on the screen.}
\end{figure}
The typical experimental setup (see~\fref{Fig:02}) consists on a mirror, a
support with a vertical rod, and a clamp holding a lamp (source of light) and a
screen horizontally~\cite{Rachna}. When there is no water in the mirror, the
centre of curvature is located at $C$, and the radius of curvature is $R=CV$.
From~\eref{Eq:f}, the focal length of the concave mirror is $f=FV=R/2$. When the
mirror is filled with water, the apparent centre of curvature $C'$ moves down,
and the apparent radius of curvature becomes $R'=C'V$. The new focal length is
given by $f'=F'V=R'/2$.

To obtain the refractive index of water $n_{\mathrm{w}}$, the clamp that holds
the lamp and the screen are moved to position $C$ until a sharp image of the
lamp is formed by the empty mirror. The radius of curvature $R$ is then
measured. Next, the mirror is filled with a thin layer of water, and the lamp
and screen moved down until a new sharp image of the lamp is formed on the
screen. This position corresponds to the apparent centre of curvature $C'$. The
new radius of curvature $R'$ is then measured. The refractive index can be
obtained by using the equation
\begin{equation}\label{Eq:RR01}
  n_{\mathrm{w}} = \frac{R}{R'}.
\end{equation}

\section*{The Snell's law approach}

\begin{figure}[h]
\centering
\includegraphics[clip=true,scale=0.75]{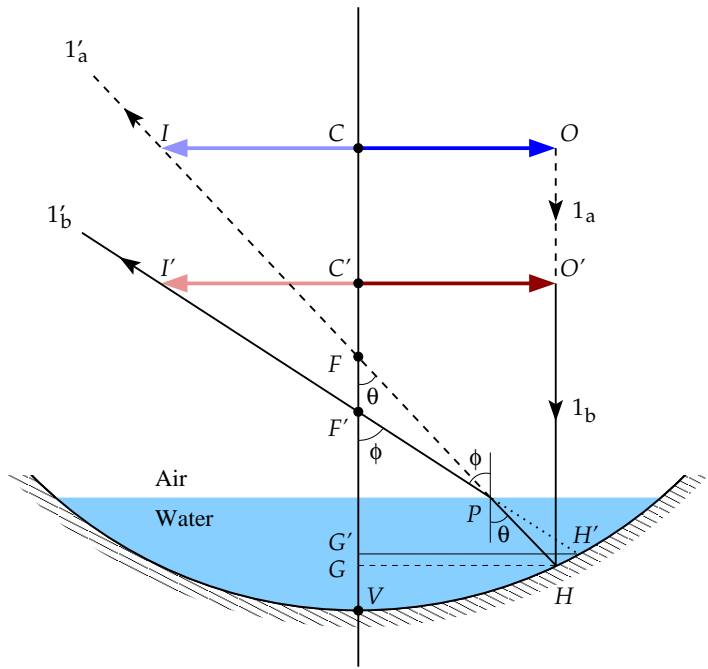}
\caption{\label{Fig:03}Formation of the images at the positions of the actual
and apparent centres of curvature.}
\end{figure}

When the point object $O$ is placed at the centre of curvature $C$, and the
mirror is empty (no water has been poured in), we may use the same graphical
ray-tracing methods of~\fref{Fig:01} to locate the conjugate image point $I$.
Ray $1_{\mathrm{a}}$ leaves point $O$ parallel to the optical axis, strikes the
mirror at point $H$ and reflects as ray $1'_{\mathrm{a}}$. This ray intersects
the principal axis at focal point $F$. The focal length is then $f=FV$. The
conjugate image point $I$ is formed at the centre of curvature $C$, as
illustrated in~\fref{Fig:03}. The image is real, inverted and has the same size
of the object.

Now, when the mirror is filled with a thin layer of water, the magnitude of its
focal length decreases, and the apparent centre of curvature $C'$ gets closer to
the vertex $V$ of the mirror. Ray $1_{\mathrm{a}}$ leaves object point $O$
parallel to the optical axis, strikes the mirror at point $H$ and then reflects.
The reflected ray refracts from water into air bending away from the surface
normal at the point of incidence $P$, for the refractive index of water
$n_{\mathrm{w}}$ is bigger than that of air $n_{\mathrm{a}}$. The refracted ray
$1'_{\mathrm{b}}$ intersects the optical axis at the new focal point $F'$. The
new focal length becomes $f' = F'V$. If the object is placed at the apparent
centre of curvature $C'$, and labelled object point $O'$, the conjugate image
point $I'$ is found at the apparent centre of curvature $C'$. The image is real,
inverted, and has the same size of the object (see \fref{Fig:03}).

Using Snell's law at surface point $P$, we get
\begin{equation}\label{Eq:Snell}
  n_{\mathrm{w}} \sin\theta = n_{\mathrm{a}} \sin\phi,
\end{equation}
where $n_{\mathrm{w}}$ and $n_{\mathrm{a}}$ are the refractive indices of water
and air, respectively. From this point on, we will assume $n_{\mathrm{a}}\approx
1.00$.

\Fref{Fig:03} shows that from triangles $FHG$ and $F'H'G'$
\numparts
\begin{eqnarray}
 HG =& FH\,\sin\theta     \label{Eq:HG01} \\
 H'G' =& F'H'\,\sin\phi.  \label{Eq:HG02}
\end{eqnarray}
\endnumparts
The backward extension of the refracted ray $1'_{\mathrm{b}}$ strikes the mirror
at point $H'$. If we assume that the focal lengths $F$ and $F'$ are large
compared to the thickness of the water layer in the mirror, points $G$ and $G'$,
as well as, $H$ and $H'$ are very close to each other. Therefore, we can make
the following approximations for the geometrical lengths:
\begin{eqnarray}\label{Eq:HG03}
\eqalign{
  H'G' \approx& HG    \\
  F'H' \approx& F'H.
}
\end{eqnarray}
Physically, points $H$ and $H'$, as well as $G$ and $G'$ are not coinciding but
are located nearby. Using~\eref{Eq:HG03} in~\eref{Eq:HG02}, we may write
\begin{equation}\label{Eq:HG04}
  HG = F'H\,\sin\phi.
\end{equation}
Since the left hand sides of~\eref{Eq:HG01} and~\eref{Eq:HG02} are the same,
we may equate both equations to obtain
\begin{equation}\label{Eq:sines}
  FH\,\sin\theta = F'H\,\sin\phi.
\end{equation}
Using~\eref{Eq:Snell} in~\eref{Eq:sines}, we obtain
\begin{equation}\label{Eq:nw01}
  \frac{FH}{F'H} = \frac{\sin\phi}{\sin\theta} = n_{\mathrm{w}}.
\end{equation}

If the angles $\theta$ and $\phi$ are small, then this angles can be replaced by
their tangents (small-angle approximation). Furthermore, the distance $GV$
in~\fref{Fig:03}, the \textit{sagittal depth} of the surface~\cite{Blaker}, is
also small, and we may neglect it. Therefore, we can write
\begin{equation}\label{Eq:nw02}
  n_{\mathrm{w}} = \frac{\tan\phi}{\tan\theta} = \frac{f}{f'}.
\end{equation}
Finally, using~\eref{Eq:f}, we can express the refractive index of water in
terms of the actual and apparent radii of curvature as follows
\begin{equation}\label{Eq:RR02}
  n_{\mathrm{w}} = \frac{R}{R'}.
\end{equation}

\section*{The silvered lens analogue}

An expression for the refractive index of the liquid may be obtained also from
the spherical mirror and thin-lens formulas
\numparts
\begin{eqnarray}
  \frac{1}{s} + \frac{1}{s'} = - \frac{2}{r} \label{Eq:mirror01}  \\
  \frac{n_1}{s} + \frac{n_2}{s'} = \frac{n_2 - n_1}{r}, \label{Eq:lens01}
\end{eqnarray}
\endnumparts
where $s$ is the object distance, $s'$ is the image distance, and $r$ is the
radius of curvature; $n_1$ and $n_2$ are the refractive indices of glass
(liquid) and air, respectively.

\begin{figure}[h]
\centering
\includegraphics[clip=true,scale=0.75]{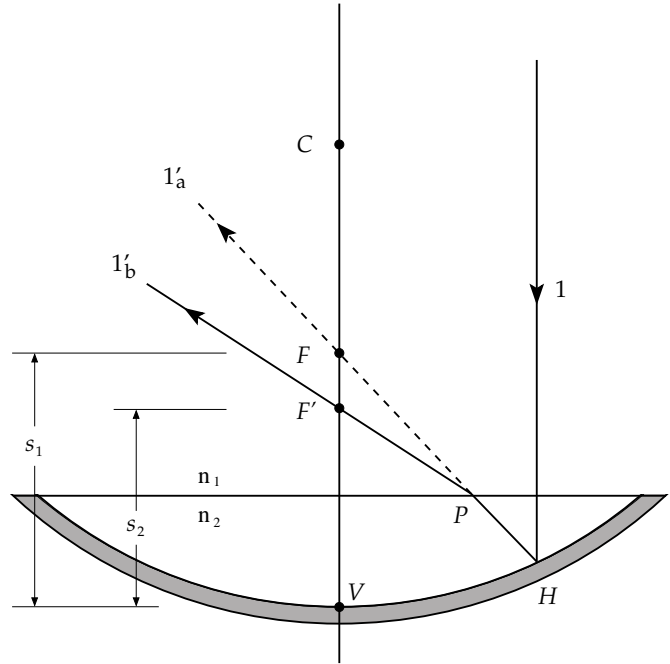}
\caption{\label{Fig:04}Schematic of the silvered lens analogue.}
\end{figure}
Consider a thin, plano-convex lens of radius of curvature $R=CV$ and silvered on
its curved side, as illustrated in~\fref{Fig:04}. For an object at infinity ($s
\rightarrow \infty$), incident rays are parallel to the optical axis. After
reflection at the curved face, we may use~\eref{Eq:mirror01} to get
\begin{equation}\label{Eq:mirror02}
  \frac{1}{\infty} + \frac{1}{s_1} = - \frac{2}{R}.
\end{equation}
Thus, $s_1 = -R/2 = f$.

Now, after refraction at the plane face
($r \rightarrow \infty$), and using~\eref{Eq:lens01} with $s=s_1$, $s'=s_2$,
$n_1=n_{\mathrm{w}}$, and $n_2=1$, we obtain
\begin{equation}\label{Eq:lens02}
  \frac{n_{\mathrm{w}}}{-R/2} + \frac{1}{s_2} = 0.
\end{equation}
Then, $s_2 = R/2n = f'$ (see~~\fref{Fig:04}).

Finally, taking the ratio of the two focal
lengths, we get
\begin{equation}\label{Eq:nw03}
  \frac{f}{f'} = \frac{R/2}{R/2n_{\mathrm{w}}} = n_{\mathrm{w}},
\end{equation}
which is in accordance with~\eref{Eq:nw02} (see reference~\cite{Pedrotti}).

\section*{Conclusions}

The experiment described in this paper gives a simple and effective method of
measuring the refractive index of water (or any transparent liquid) using a
spherical concave mirror.

A valuable pedagogical consideration of this method is that, students may
observe that a real image of the object is formed at the centre of curvature of
the empty mirror when the object is located at the same point. Similarly, a real
image is formed at the apparent centre of curvature when the mirror is filled
with water, if the object is located at the same place. This provides an easy
way of finding the refractive index of the liquid by measuring the ratio between
the actual and apparent radii of curvature. In addition, by using Snell's law
and reflection of rays in a concave spherical mirror, they should notice that
the images are not only real, but inverted, and have the same size of the
object.

As shown by the Snell's law approach, with the concave mirror method we can get
an expression for the refractive index of a liquid using merely Snell's law and
the small angle approximation (paraxial approximation), and avoiding
complications introduced by the thin-lens equation. However, in the silvered
mirror analogue, we use the spherical mirror and thin-lens equations to verify
the validity of the refractive index equation obtained using the Snell's law.

This experiment is suitable for the undergraduate physics laboratory. It is easy
to setup and perform. It has been carried out successfully in the lab providing
accurate numerical values of the refractive index of water and other transparent
liquids.

\ack
Author (AJ) gratefully acknowledges funding support from RCSA.


\end{document}